\documentclass[review]{elsarticle}
\usepackage{amsmath}
\usepackage{amsfonts}

\usepackage{graphicx}  

\usepackage{mathrsfs}
\usepackage{enumerate}
\usepackage{tikz}
\usetikzlibrary{decorations.pathmorphing,patterns}
\usepackage{tikz-cd}

\usepackage{subfigure}
\usepackage{soul}
\usepackage{times} 
\usepackage{amssymb}  
\usepackage[english]{babel}

\newcommand{\R}{\mathbb{R}}

\newcounter{met}

\newtheorem{theorem}{Theorem}[section]
\newtheorem{corollary}{Corollary}
\newtheorem{lemma}[theorem]{Lemma}
\newtheorem{proposition}{Proposition}

\newtheorem{example}{Example}

\begin{document}

\begin{frontmatter}

\title{
Variational point-obstacle avoidance on Riemannian manifolds\tnoteref{mytitlenote}}
\tnotetext[mytitlenote]{The research of A. Bloch was supported by NSF grants DMS-1207693, DMS-1613819 and ASFOR. The research of M. Camarinha was partially supported by the Centre for Mathematics of the University of Coimbra -- UID/MAT/00324/2019, funded by the Portuguese Government through FCT/MEC and co-funded by the European Regional
Development Fund through the Partnership Agreement PT2020. L. Colombo was supported by This work was partially supported by I-Link Project (Ref: linkA20079) from CSIC; Ministerio de
Economia, Industria y Competitividad (MINEICO, Spain) under grant MTM2016-76702-P; `Severo
Ochoa Programme for Centres of Excellence" in R$\&$D (SEV-2015-0554).The project that gave rise to these results received the support of a fellowship from ``la Caixa' Foundation (ID 100010434). The
fellowship code is LCF/BQ/PI19/11690016.}

\author{Anthony M. Bloch\fnref{1}}
\address{Department of Mathematics, University of Michigan}
\address{530 Church St. Ann Arbor, 48109, Michigan, USA}
\fntext[1]{Email adress: abloch@umich.edu.}

\author{Margarida Camarinha \fnref{2}}
\address{Centre for Mathematics of the University of Coimbra, Department of Mathematics}
\address{University of Coimbra, 3001-501 Coimbra, Portugal}
\fntext[2]{Email adress: mmlsc@mat.uc.pt.}

\author{Leonardo J. Colombo\fnref{3}}
\address{Instituto de Ciencias Matem\'aticas, Consejo Superior de Investigaciones Cient\'ificas}
\address{Calle Nicol\'as Cabrera 13-15, Cantoblanco, 28049, Madrid, Spain}
\fntext[3]{Email adress: leo.colombo@icmat.es.}


\begin{abstract}
In this letter we study variational obstacle avoidance problems on complete Riemannian manifolds. The problem consists of minimizing an energy functional depending on the velocity, covariant acceleration and a repulsive potential function used to avoid a static obstacle on the manifold, among a set of admissible curves. We derive the dynamical equations for extrema of the variational problem, in particular on compact connected Lie groups and Riemannian symmetric spaces. Numerical examples are presented to illustrate the proposed method.
\end{abstract}

\end{frontmatter}

\section{Introduction}
Many problems in physics, engineering and related disciplines can be formulated as variational problems. Sometimes the solution we seek has to satisfy some nonlinear constraints or avoid static or moving obstacles in the space of configurations of a given system. Riemannian manifolds are, in many cases, the suitable configuration spaces to model these problems. Variational problems on Riemannian manifolds have been extensively studied in the last decades for applications ranging from trajectory planning in aerospace engineering \cite{BlochHussein}, \cite{Jackson}, interpolation of data in medical images and pattern recognition \cite{Hinkle} to parametric regression of data for computer vision problems \cite{Hong}. A basic reference for  variational theory on Riemannian manifolds is the book of Milnor \cite{Milnor}. This key procedure, which is Lagrangian in nature, is to study the characterization  of critical paths of an action functional over a set of admissible curves.

Since then, a number of papers have been devoted to the generalization of this variational theory in many other contexts: interpolation problems \cite{BlCaCoIJC}, collision avoidance of multiple agents \cite{mishal} and quantum splines interpolation \cite{ligia}, among others. There are various treatments of obstacle avoidance in different contexts, nevertheless, to the best of our knowledge, the point-obstacle avoidance  problem, that is, the problem of creating feasible, safe paths that avoid a prescribed point-obstacle while minimizing some quantity such as energy or time in the Riemannian manifold setting has not been widely discussed in the literature.

We studied trajectory planning schemes with obstacle avoidance on Riemannian manifolds from the variational point of view in our previous work \cite{BlCaCoCDC}. Inspired by the goal of gaining a better understanding of trajectories which minimize a weighted combination of the covariant acceleration and  the velocity of the system in the presence of a repulsive potential which is  used to avoid a static circular obstacle, in \cite{BlCaCoIJC} we extended the problem to the trajectories that also interpolate some points on the manifold. The present work goes one step further and considers variational obstacle avoidance problems on complete Riemannian manifolds with the obstacle being a specified configuration represented by an element in the manifold. The aim is to study necessary optimality conditions for the problem for different systems on Riemannian  Lie groups and symmetric spaces.

Specifically, the problem studied in this work consists of finding  variational trajectories surrounding a given obstacle, among a set of admissible curves, which minimize an energy
functional that depends on the velocity and covariant acceleration. An artificial potential function is used to prevent the trajectory to cross a given point-obstacle. To solve the problem, we employ techniques from the calculus of variations on Riemannian manifolds, taking into account that the problem under study can be seen as a higher order variational problem \cite{BlCr}, \cite{MargaridaThesis}, \cite{Cophd}, \cite{CoMdM}, \cite{CoFeMdD}, \cite{BlochHussein}.

The main contribution is to provide necessary optimality conditions for the obstacle avoidance problem  on a Riemannian manifold (i.e., a nonlinear space), based on differential equations on a vector space (i.e., a linear space).

This procedure is possible due to the bi-invariance of the Riemannian metric on the Lie group, which allows us to left translate the higher-order covariant derivatives of the trajectory to the Lie algebra.  The potential function is expressed in terms of the exponential map, defined by the geodesic connecting the  configuration of the system with  the obstacle configuration, and its gradient can also be left translated to the identity.  One advantage of this method consists in the fact that, if we are dealing with the problem in a symmetric space, then we can lift the trajectories to the Lie group acting on it, solve the equations there, and project back the trajectories to the symmetric space. The assumption of having complete manifolds is essential to guarantee that every geodesic in the symmetric space is the projection of a horizontal geodesic in the Lie group, thus allowing one to study the problem in the Lie group. From this point of view, the problem studied in this work extends recent developments for cubics in tension on symmetric spaces presented in \cite{ZN}, \cite{ZN1}.

The structure of the paper is as follows. We start by introducing the geometric framework  on a Riemannian manifold that is necessary to study the variational problem in Sec. II. Next, in Sec. III, we introduce variational point-obstacle avoidance problems on complete Riemannian manifolds and we derive necessary optimality conditions. A special emphasis is given to the case of compact and connected Lie groups with the illustrative example of the rigid body in SO(3). Finally in Section V we analyse the problem on Riemannian symmetric spaces lifting the equations to the Lie group acting on the manifold. Numerical examples on the sphere $S^2$ are presented to illustrate the proposed method. 

\section{Preliminaries on Riemannian Geometry }
Let $M$ be an $n$-dimensional  \textit{Riemannian
manifold} with \textit{Riemannian metric} denoted by
$\langle\cdot,\cdot\rangle:T_xM\times T_xM\to\mathbb{R}$ at each point $x\in M$, where $T_xM$ is
the \textit{tangent space} of $M$ at $x$. The length of a tangent vector is determined by its norm,
$||v_x||=\langle v_x,v_x\rangle^{1/2}$ with $v_x\in T_xM$. A \textit{Riemannian connection} $\nabla$ on $M$ is a map that assigns to any two smooth vector fields $X$ and $Y$ on $M$ a new vector field, $\nabla_{X}Y$. For the properties of $\nabla$, we refer the reader to \cite{Boothby, bookBullo,Milnor}.  The operator
$\nabla_{X}$, which assigns to every vector field $Y$ the vector
field $\nabla_{X}Y$, is called the \textit{covariant derivative of
$Y$ with respect to $X$}.

Given vector fields $X$, $Y$
and $Z$ on $M$, the vector field $R(X,Y)Z$ given by \begin{equation}\label{eq:CurvatureTensorDefinition}
R(X,Y)Z=\nabla_{X}\nabla_{Y}Z-\nabla_{Y}\nabla_{X}Z-\nabla_{[X,Y]}Z
\end{equation}  is called the \textit{curvature tensor} of $M$. $R$ is trilinear in $X$, $Y$ and $Z$, and a tensor of type $(1,3)$.

Consider a vector field $W$  along a curve $x$ on $M$. The $s$th-order covariant  derivative along $x$  of $W$ is denoted by $\displaystyle{\frac{D^{s}W}{dt^{s}}}$, $s\geq 1$. We also denote by $\displaystyle{\frac{D^{s+1}x}{dt^{s+1}}}$ the $s$th-order covariant derivative along $x$ of the velocity vector field of $x$, $s\geq 1$.

A vector field $X$ along a piecewise smooth curve $x$ in $M$ is said to be \textit{parallel along $x$} if $\displaystyle{\frac{DX}{dt}\equiv 0}$. If $x_0=x(0)$ is the initial point of the curve $x$ and $Y\in T_{x_0}M$ is an arbitrary tangent vector to $M$ at $x_0$, then there exists a unique parallel vector field $X$ along $x$ having the value  $Y$ at $x_0$. When the velocity vector field of a curve $x$ is parallel, the curve $x$ is called a geodesic.

We assume here that $M$ is \textit{complete}, which implies that any two points $p$ and $q$ in $M$ can be connected by a shortest arc  $\gamma_{p,q}$. In such a context the Riemannian distance between two points in $M$, $d:M\times M\to\mathbb{R}$ can be defined by
$$d^2(p,q)=\int_{0}^{1}\Big{\|}\frac{d \gamma_{p,q}}{d s}(s)\Big{\|}^2\, ds.$$
Additionally, if we assume that the points $p$ and $q$ belong to a convex open  ball $\mathcal{B}$, the Riemannian exponential map $\mbox{exp}_q$ is a diffeomorphism in $\mathcal{B}$ and we can  write the Riemannian distance by means of the Riemannian exponential on $M$ as
$$d(q,p)=\|\mbox{exp}_q^{-1}p\|.$$

%
%
%

\section{Variational obstacle avoidance problem on complete Riemannian manifolds}

\subsection{Problem formulation and dynamical equations}

Let $M$ be a complete Riemannian manifold, $T$, $\sigma$ and $\tau$  be positive real numbers, $(p_0,v_0)$, $(p_T,v_T)$ points in $TM$  and $S$ a regular submanifold of $M$. Consider the set $\Omega$ of \textit{admissible curves}, all  ${\mathcal C}^{1}$ piecewise smooth curves  on $M$, $x:[0,T]\rightarrow M$, verifying the boundary conditions \begin{equation}\label{boundaryc}
x(0)=p_0, \quad  x(T)=p_T,\quad \frac{dx}{dt}(0)=v_0, \quad \frac{dx}{dt}(T)=v_T,\end{equation} and define the functional $J$ on $\Omega$ given by
\begin{equation}\label{3.2}
J(x)=\int_{0}^{T}\frac{1}{2}\left(\Big{\|}\frac{D^2x}{dt^2}(t)\Big{\|}^2+
 \sigma \Big{\|}\frac{ dx}{dt}(t)\Big{\|}^2+V(x(t))\right)dt.
\end{equation}

\noindent The functional \eqref{3.2} is  given by a weighted combination of the velocity and  covariant acceleration of the curve $x$ regulated by the
parameter $\sigma$, together with an artificial potential function $V:M\to\mathbb{R}$ used to ensure collision avoidance with an static obstacle which is given by a configuration $q$ on $M$.

 The function $V$ is assumed to be at least $C^{2}$ and associated with a fictitious force inducing a repulsion from $q$, defined as the inverse value of a distance function specified by the Riemannian exponential. Then, $V$  goes to infinity  near $q$ and decays to zero at some positive level set far away from the obstacle $q$. This ensures that  trajectories given by solutions of the variational problem do not intersect $q$.  The use of artificial potential functions to avoid collision was introduced in Khatib (see \cite{Khatib1986} and references therein) and further studied for example by Koditschek and Rimon \cite{K88}, \cite{Koditschek1990}.

For the class of admissible curves $x$, we introduce the ${\mathcal C}^1$-piecewise smooth \textit{one-parameter admissible variation} of  $x$ as $
\alpha : (-\epsilon , \epsilon ) \times [0,T]  \rightarrow  M ;(r,t) \mapsto \alpha (r,t)=\alpha_r(t)$
verifying $\alpha_0(t)=x(t)$ and $\alpha_r\in\Omega$, for each $r\in (-\epsilon , \epsilon )$.
The \textit{variational vector field}  associated to a one-parameter admissible variation $\alpha$ is a ${\mathcal C}^{1}$-piecewise smooth vector field along $x$ defined by $$X(t)=\frac{D}{\partial r}\Big{|}_{r=0}\alpha(r,t)\in T_{x}\Omega,$$
verifying  the boundary conditions
\begin{equation}\label{3.6}
X(0)=0,\quad   X(T)=0,\quad
\frac{DX}{dt}(0)=0,\quad   \frac{DX}{dt}(T)=0.\end{equation}

The admissible set  $\Omega$ admits an infinite dimensional Hilbert manifold structure (see \cite{Giambo} and references therein) and its tangent space  $T_{x}\Omega$ at $x$ can be identified with the
 vector space   of all ${\mathcal C}^{1}$ piecewise smooth vector
fields $X$ along $x$  verifying the boundary conditions (\ref{3.6}).

\textbf{Problem}: The \textit{variational obstacle avoidance problem} consists of minimizing the functional $J$ among $\Omega$ satisfying the boundary condition \eqref{boundaryc}.

In \cite{BlCaCoCDC} we proved the following result as a solution of the previous problem:

\begin{theorem} \label{t3.22}
A necessary condition for $x$  to be a minimizer of the functional \eqref{3.2} over the class $\Omega$ satisfying the boundary condition \eqref{boundaryc} is that, $x$ is smooth on [0,T], and verifies the following equation
 \begin{equation} \label{3.8}
\frac{D^{4}x}{dt^{4}}+R\left(\frac{D^2x}{dt^2},\frac{dx}{dt}\right)\frac{dx}{dt}- \sigma
\frac{D^2x}{dt^2}+ \frac{1}{2}\mbox{grad }V(x) \equiv 0.
\end{equation}

\end{theorem}
We consider the repulsive potential function defined by $\displaystyle{V(x)=\frac{\tau}{d^2(q,x)}}$, $x\in M$. Such a potential function gives tractable formulas for the gradient of $V$ in terms of the exponential map as we will see in Lemma \ref{prop}(see \cite{Karcher77} for more details in the subject).

In this paper we assume that the points $x_0,x_T\in M$ are sufficiently close to guarantee the exponential map is global diffeomorphism, which means that we restrict our analysis to an convex open neighborhood of the obstacle containing $x_0$ and $x_T$.

\begin{lemma}\label{prop}
If $B_q$ is a convex open ball containing $q$ and $V$ is the function defined  by $\displaystyle{V(p)=\frac{\tau}{d^2(q,p)}}$ in $B_q$, then its gradient can be written in the form $$\hbox{grad }V(p)=\frac{\tau}{d^4(q,p)}\exp_{p}^{-1}q.$$
\end{lemma}

\textit{Proof:}

If we consider a map $\alpha: r\to \alpha(r)$ verifying $\alpha(0)=x$  and  the family of geodesics from $q$ to $\alpha(r)$ is given by
$\gamma(s,r)=\mbox{exp}_q(s\, \mbox{exp}_q^{-1}\alpha(r)),$ then we have
$$\displaystyle \frac{d}{d r}d^2(q,\alpha)=\Big{\langle}\frac{\partial \gamma}{\partial r },\frac{\partial \gamma}{\partial
s}\Big{|}_{s=1} \Big{\rangle}=-\Big{\langle}\frac{d \alpha}{dr },\mbox{exp}_{\alpha}^{-1}q\Big{\rangle}.$$
Therefore, the gradient vector field of $V$ is given by
$\displaystyle{\mbox{grad }V(x)=\frac{\tau}{d^4(q,p)}\exp_{p}^{-1}q}$. \hfill$\square$

As a consequence of Theorem \ref{t3.22} and Lemma \ref{prop} we have the following result for the obstacle avoidance problem on a complete Riemannian manifold.

\begin{corollary}\label{t3.3} A necessary condition for $x$  to be a minimizer of the functional \eqref{3.2} over the class $\Omega$ satisfying the boundary condition \eqref{boundaryc} is that, $x\in M$ is smooth on $[0,T]$, and verifies the following equations
\begin{equation}\label{eqwithexp}
\frac{D^{4}x}{dt^{4}}+R\left(\frac{D^2x}{dt^2},\frac{dx}{dt}\right)\frac{dx}{dt}- \sigma
\frac{D^2x}{dt^2}+\frac{1}{2}\frac{\tau}{d^4(q,x)}\exp_x^{-1}q \equiv 0.
\end{equation}

\end{corollary}

\subsection{Variational obstacle avoidance problem on compact and connected Lie groups}

Let $G$ be a compact and connected Lie group endowed with a bi-invariant Riemannian metric $\langle \cdot , \cdot \rangle$ and $\mathfrak{g}$ its Lie algebra. The following result from \cite{Nomizu} provides a formula for the covariant derivative $\nabla$ and the curvature tensor $R$ in terms of the Lie algebra structure.

\begin{lemma}\label{thformulas}\hspace{.5cm}\begin{itemize}
\item[(i)] Every compact and connected Lie group admits a left and right invariant metric $\langle\cdot,\cdot\rangle$.
\item[(ii)] If $\nabla$ denotes the corresponding Levi-Civita  connection induced by the metric $\langle\cdot,\cdot\rangle$ and $X,Y$ and $Z$ are left-invariant vector fields on $G$ then $$\nabla_{X}{Y}=\frac{1}{2}[X,Y] \mbox{ and } R(X,Y)Z=-\frac{1}{4}[[X,Y],Z].$$
\end{itemize}
\end{lemma}

This lemma guarantees that the connection $\nabla$ is completely determined by its restriction to $\mathfrak{g}$ via left-translations. This restriction, denoted by $\stackrel{\mathfrak{g}}{\nabla}:\mathfrak{g}\times\mathfrak{g}\to\mathfrak{g}$,  is naturally given by $\stackrel{\mathfrak{g}}{\nabla}_wu= \frac 12 [w,u]$ (see \cite{bookBullo} p. 271).
Indeed, if $u,w\in\mathfrak{g}$ we have $\nabla_{w_L}u_L=(\stackrel{\mathfrak{g}}{\nabla}_wu)_L$, where $u_{L}$ denotes the left-invariant vector field  associated  to $u$.


Let $x:I\subset\mathbb{R}\to G$ be a smooth curve on $G$. The \textit{body velocity} of $x$ is the curve $v:I\subset\mathbb{R}\to\mathfrak{g}$ defined by $\displaystyle{v(t)=T_{x(t)}L_{x(t)^{-1}}\left(\frac{dx}{dt}(t)\right)}$, where $L_g:G\to G$ denotes the left-translation map by $g$.

Let $\{e_1,\ldots,e_n\}$ be a basis of  $\mathfrak{g}$. Consider the body velocity of $x$ on the given basis, defined by $\displaystyle{v=\sum_{i=1}^n v_i e_i}$.
To write the equations determining necessary conditions for extremal, we use the following formulas,  where $\stackrel{\mathfrak{g}}{\nabla}_wv=\displaystyle{\sum_{i,j=1}^n w_i v_j\nabla_{e_j}e_i}$ and $e$ denotes the identity element on $G$ (see for instance \cite{BlCaCoCDC}).
\begin{align*}
&\frac{dx}{dt}=T_eL_{x}v,\quad \frac{D^{2}x}{dt^{2}}=T_eL_{x}\Big(v^{\prime}+\stackrel{\mathfrak{g}}{\nabla}_vv\Big),\\
&\frac{D^{3}x}{dt^{3}}=T_eL_{x}\Big(v^{\prime \prime}+\stackrel{\mathfrak{g}}{\nabla}_{v^{\prime}}v+2 \stackrel{\mathfrak{g}}{\nabla}_vv^{\prime}+\stackrel{\mathfrak{g}}{\nabla}_v\stackrel{\mathfrak{g}}{\nabla}_vv\Big),\\
&\frac{D^{4}x}{dt^{4}}=T_eL_{x}\left(v'''+\stackrel{\mathfrak{g}}{\nabla}_{v''}v+3\stackrel{\mathfrak{g}}{\nabla}_{v^{\prime}}v^{\prime}+3 \stackrel{\mathfrak{g}}{\nabla}_vv''+\stackrel{\mathfrak{g}}{\nabla}_{v^{\prime}}\stackrel{\mathfrak{g}}{\nabla}_vv+2 \stackrel{\mathfrak{g}}{\nabla}_v\stackrel{\mathfrak{g}}{\nabla}_{v^{\prime}}v+3\stackrel{\mathfrak{g}}{\nabla}_v^2v^{\prime}+\stackrel{\mathfrak{g}}{\nabla}^{3}_vv\right),\\
&R\left(\frac{D^{2}x}{dt^{2}},\frac{dx}{dt}\right)\frac{dx}{dt}=T_eL_{x}\left(\mathfrak{R}(v^{\prime},v)v+\mathfrak{R}(\stackrel{\mathfrak{g}}{\nabla}_vv,v)v\right),\end{align*} where $\mathfrak{R}$ denotes the curvature tensor associated with $\stackrel{\mathfrak{g}}{\nabla}$.

Using Theorem \ref{thformulas}, the previous formulas are reduced to \begin{align}
&\frac{D^{2}x}{dt^{2}}=T_eL_{x}v^{\prime},\label{eqq2-2}\\
&\frac{D^{3}x}{dt^{3}}=T_eL_{x}\Big(v^{\prime \prime}+\frac{1}{2}[v, v^{\prime}]\Big),\label{eqq3-2}\\
&\frac{D^{4}x}{dt^{4}}=T_eL_{x}\Big(v'''+[v,v'']+\frac 14[[v^{\prime},v],v]\Big),\label{eqq4-2}\\
&R\left(\frac{D^{2}x}{dt^{2}},\frac{dx}{dt}\right)\frac{dx}{dt}=-\frac 14T_eL_{x}\Big([[v^{\prime},v],v]\Big)\label{eqq5-1}.\end{align}

\begin{lemma}\label{propreduceeq}
Let $G$ be a compact and connected Lie group, then for all $x,y\in G$ the following identity holds $$\exp_{y}^{-1}(x)=T_{e}L_{y}\exp_{e}^{-1}(y^{-1}x)$$
\end{lemma}

\textit{Proof:} Let $\gamma$ be a geodesic starting at $\gamma(0)=e$ and finishing at $\gamma(1)=y^{-1}x$ with $\gamma'(0)=\exp_{e}^{-1}(y^{-1}x)$. The curve $\beta(t)=L_{y}\circ\gamma(t)$ is a curve such that $\beta(0)=y$ and $\beta(1)=x$ with $\beta'(0)=T_{e}L_{y}\exp_{e}^{-1}(y^{-1}x)$.

Given that $\nabla$ is left-invariant, $\beta$ is a geodesic, and therefore $T_{e}L_{y}\exp_{e}^{-1}(y^{-1}x)=\exp_{y}^{-1}(x)$.\hfill$\square$

\vspace{.2cm}

Using \eqref{eqq2-2}-\eqref{eqq5-1} and Lemma \eqref{propreduceeq} the equations \eqref{eqwithexp} for  the variational obstacle avoidance problem on a connected and compact Lie group are given as follows.

\begin{proposition}\label{Theorem3.5}
Let $x$ be a curve on a connected and compact Lie group $G$ with body velocity $v$ with respect to the basis $\{e_1,\ldots,e_n\}$ in $\mathfrak{g}$. A necessary condition for $x$  to be a minimizer of the functional \eqref{3.2} over the class of curves $\Omega$ satisfying the boundary conditions \eqref{boundaryc} is that, $x$ is smooth on $[0,T]$, and the curve $v$ in $\mathfrak{g}$ verifies the following equation
 \begin{equation} \label{completeG}
v'''-\sigma v'+[v,v'']+\frac{\tau}{2||\exp_{e}^{-1}(x^{-1}q)||^4}\exp_{e}^{-1}(x^{-1}q) \equiv 0.
\end{equation}
\end{proposition}

\begin{example}

This example is motivated by the fact that obstacle avoidance problems defined on the special orthogonal group $SO(3)$ are often used to model avoidance of certain orientations of the rigid body. This is for instance the case for planning motion of an optical instrument where avoiding pointing at a certain light source is crucial.

We consider the variational obstacle avoidance problem on the Lie group $SO(3)$. The Lie algebra $\mathfrak{so}(3)$ is given by the set of $3\times 3$ skew-symmetric matrices.

Denote by $t\to R(t)$ a curve on $SO(3)$. The columns of the
matrix $R(t)$ represent the directions of the principal axis of the
 body reference system at time $t$ with respect to some fixed reference system.

 It is well known that $\mathfrak{so}(3)\simeq \R^3$, where the Lie bracket of matrices is identified with the cross product. This Lie algebra isomorphism  is the hat map $\hat{\cdot}:\mathbb{R}^{3}\to\mathfrak{so}(3)$ that assigns a matrix $A\in\mathfrak{so}(3)$, that is, a skew-symmetric matrix of the form $A=\left(
               \begin{array}{ccc}
                 0 &-a_3  & a_2 \\
                 a_3 & 0 & -a_1 \\
                 -a_2 & a_1 & 0 \\
               \end{array}
             \right)$  to the vector $a=(a_1,a_2,a_3)\in \mathbb{R}^3$. The matrix $A$ can be denoted by  $\widehat{a}$. We endow $SO(3)$ with the bi-invariant metric  $\langle\cdot , \cdot \rangle$  corresponding to the usual inner product in $\R^3$ via the hat isomorphism.
By Lemma \ref{thformulas}, the Levi-Civita connection $\nabla$ induced by $\langle\cdot , \cdot \rangle$ is completely determined by its restriction to the Lie algebra $\mathfrak{so}(3)$ given by
$\stackrel{\mathfrak{so}(3)}{\nabla}_{v} z=\frac{1}{2}v\times z$
and the  restriction of the curvature tensor  to  $\mathfrak{so}(3)$ is defined by
$\displaystyle{\mathfrak{R}(v,z)w=-\frac 14 (v\times z)\times w}$
where $v,z,w \in \mathbb{R}^3$.

For the obstacle avoidance problem we consider the artificial potential $V:SO(3)\to\mathbb{R}$ given by \begin{equation*}\label{navfunct3}V(R)=\frac{\tau}{\|\mbox{exp}^{-1}(R^{T}Q)\|^2}
\end{equation*} with $\tau\in\mathbb{R}^{+}$ and $Q,R\in SO(3)$, with $\exp:\mathfrak{so}(3)\to SO(3)$ denoting the matrix exponential map on $SO(3)$ given by $\displaystyle{\exp\xi=e^{\xi}}$ with $\xi\in\mathfrak{so}(3)$.  The matrix exponential map is a diffeomorphism between $U=\{ \widehat{a}\in\mathfrak{so}(3): \|a\|<\pi, a\in \R^3\}$ and  $V=\{R\in SO(3): \hbox{Tr}\neq -1\}$ and  its inverse map is  the matrix logarithm map.
Using the matrix logarithm, \begin{equation}\label{eqlog}\frac{\exp^{-1}(R^{T}Q)}{\|\exp^{-1}(R^{T}Q)\|^4}=\frac{\log(R^{T}Q)}{||\log(R^{T}Q)||^{4}}.\end{equation} Denoting $\phi=\hbox{arccos}(\frac{1}{2}(\hbox{Tr}(R^{T}Q)-1)$, and using Proposition $5.7$ in \cite{bookBullo},  for $R\neq Q$\footnote{If $R=Q$, $\log(R^{T}Q)=0$ and this case is outside the problem formulations since we are in the obstacle}, $$\log(R^{T}Q)=\frac{\phi}{2\sin(\phi)}(R^{T}Q-R^{T}Q).$$  Since $\|\log(R^{T}Q)\|=\phi$ it follows that

$$\frac{\log(R^{T}Q)}{||\log(R^{T}Q)||^{4}}=\frac{1}{2\phi^3\sin(\phi)}(R^{T}Q-RQ^{T}).$$

The body velocity of the curve $R$ in $SO(3)$ is the curve $v$  in $\mathfrak{so}(3)$ verifying $R'=Rv$. Therefore, by  Proposition \ref{Theorem3.5} the minimizers for the obstacle avoidance problem on $SO(3)$ verify the equation

\begin{equation}\label{obstacleSO(3)}v'''= v''\times v + \sigma v'+\frac{\tau}{4\phi^3\sin(\phi)}(RQ^{T}-QR^{T})^{\times}.\end{equation}
together with the equation $R'=Rv$, $\sigma\in\mathbb{R}$, $\tau\in\mathbb{R}^{+}$ and the boundary conditions $R(0)=R_0$, $R(T)=R_N$, $v(0)=v_0$, $v(T)=v_N$, where $\times:\mathfrak{so}(3)\to\mathbb{R}^{3}$ denotes the inverse of the hat isomorphism $\hat{\cdot}$.

Note that in the absence of obstacles, the extremals  reduce to the cubic polynomials in tension on $SO(3)$ (see \cite{SCC00}) which equations are given by solutions of the equation $ v'''= v''\times v + \sigma v'$.
\end{example}

\section{Application to variational obstacle avoidance problem on Riemannian symmetric spaces}
Let $H:=G/K$ be a Riemannian symmetric space, where $G$ is a connected  finite-dimensional Lie group endowed with a bi-invariant Riemannian metric
and   $K$ a closed Lie subgroup of $G$.

It is well known that the canonical projection $\pi:G\to H$ is a Riemannian submersion (see \cite{Helgason} for instance), in the sense that, for all $g$ in $G$, the isomorphism $T_g\pi:(\ker T_g\pi)^{\bot} \to T_{\pi(g)}H$ preserves the inner-products defined by the Riemannian metrics on $G$ and $H$ and $T_gG$ splits naturally into two orthogonal subspaces, the vertical subspace $V_g:=\ker T_g\pi$ and the horizontal subspace $\hbox{Hor}_g=(V_g)^{\perp}:=(\ker T_g\pi)^{\bot}$. The corresponding projections of $T_gG$ onto $V_g$ and $\hbox{Hor}_g$ are denoted by $\mathcal{V}$ and $\mathcal{H}$.

In particular,  the Lie algebra $\mathfrak{g}$ of $G$ admits the decomposition
$\mathfrak{g}=\mathfrak{s}\oplus\mathfrak{m}$ where $\mathfrak{s}$ is the Lie algebra of $K$ and $\mathfrak{m}\simeq T_{\mathfrak{o}}H$, whith  $\mathfrak{0}=\pi(e)$, being $e$ the identity element on $G$. That is, $\ker T_e\pi=\mathfrak{s}$  and the horizontal subspace $(\ker T_g\pi)^{\bot}$ is $\mathfrak{m}$. Moreover, the following relations hold (see \cite{Helgason})
$$[\mathfrak{s},\mathfrak{s}]\subset\mathfrak{s},\quad[\mathfrak{m},\mathfrak{m}]\subset\mathfrak{s},\quad[\mathfrak{m},\mathfrak{s}]\subset\mathfrak{m}.$$

Using the decomposition of $T_gG$ and defining vertical and horizontal tangent vectors on $G$, it is possible to define horizontal  curves and vector fields on $H$  to $G$ , by choosing horizontal tangent vectors. We consider the horizontal curve $\tilde{x}$ on $G$ verifying equations $\displaystyle{v=T_{\tilde{x}(t)}L_{\tilde{x}^{-1}}\left(\frac{d\tilde{x}}{dt}\right)}$ and
\begin{equation}\label{eqwithexp4}
v'''+[v, [v^{\prime},v]]- \sigma v^{\prime}+\frac{1}{2}\frac{\tau}{\|\exp_e^{-1}(\tilde{x}^{-1}\tilde{q})\|^4}\exp_e^{-1}(\tilde{x}^{-1}\tilde{q})\equiv 0,
\end{equation} the latter equation being that on the subspace $\mathfrak{m}$ and with $\tilde{q}\in\pi^{-1}(q)$.

According to \cite{ZN} and \cite{Oneill}, if we project $\tilde{x}$ to $H$ by $\pi$, we obtain a curve $x$ on $H$ verifying equations \eqref{eqwithexp}, that is, if we are able to find a curve $v$, and the corresponding curve $\tilde{x}$, verifying \eqref{eqwithexp4} and $\displaystyle{\frac{d\tilde{x}}{dt}=T_{e}L_{\tilde{x}}v}$, a solution  of  \eqref{eqwithexp} can by obtained by  projecting the curve $\tilde{x}$ to $H$.

\begin{example}
Consider the symmetric space $H=S^{2}$, the two-dimensional unit sphere, where $G=SO(3)$ and $K=SO(2)$. Denoting the canonical basis of $\R^3$ by $\{e_1,e_2,e_3\}$, the group $SO(2)$ can be seen as the subgroup of $SO(3)$ leaving $e_1 \in S^2$ fixed. 

Each element $x$ of $S^2$ can be represented by an element $R$ in $SO(3)$ by the relation $x=Re_1$ and the projection $\pi:SO(3)\to S^{2}$ is given by $\pi(R)=Re_1$. Note that the Lie algebra decomposition
$\mathfrak{so}(3)=\mathfrak{s}\oplus\mathfrak{m}$ corresponds to $\R^3=\widehat{\mathfrak{s}}\oplus\widehat{\mathfrak{m}}$ via the hat isomorphism, with $\mathfrak{s}=\hbox{span}\{e_1\}$  and $\widehat{\mathfrak{m}}=\hbox{span}\{e_2,e_3\}$.

Let $q$ be the obstacle in $S^{2}$. To obtain the extremals for the obstacle avoidance problem it is enough to solve  the following differential equations in $SO(3)\times \widehat{\mathfrak{m}}
$

\begin{equation}\label{eqwithexp4}
v'''=(v^{\prime}\times v)\times v+\sigma v^{\prime}+\frac{\tau}{4\phi^3\sin(\phi)}(RQ^{T}-QR^{T})^{\times}
\end{equation} and $R'=Rv$, with $Q\in SO(3)$ verifying the condition $\pi(Q)=q$. Next we project the solution $R$ to $S^2$.

\textbf{Simulation results:} We now show some simulation results demonstrating applicability of the proposed method in the obstacle avoidance problem on the sphere. In all simulations we employ an Euler method with step size $h=0.001$. We consider $R(0)$ to be  the $(3\times 3)$ identity matrix for all the simulations.

\begin{enumerate}\item[(1)] \textit{Obstacles along a geodesic}: We first consider a situation with $\sigma=0$. In the absence of an obstacle, the solution of equation \eqref{eqwithexp4} is a geodesic. We consider initial values $v(0)=(0,0,1)$, $v'(0)=v''(0)=(0,0,0)$ and display the result in Figure \ref{fig1}. We chose a random value $q$ which is the obstacle along the geodesic, and next we found a representative for $q$ in $SO(3)$, denoted $\tilde{Q}$. Note that we have the freedom of choosing one angle in $SO(3)$ since the sphere only provides two of the three Euler angles describing an element of $SO(3)$ (i.e., we have infinitely many choices of such a value). We choose $\pi/4$. We solve the equations on $SO(3)$ and then project back to the sphere $S^2$. In Figure \ref{fig1} we display, with $\tau=1$ the solution on $S^2$.
\begin{figure}[h!]
\includegraphics[width=.49\linewidth]{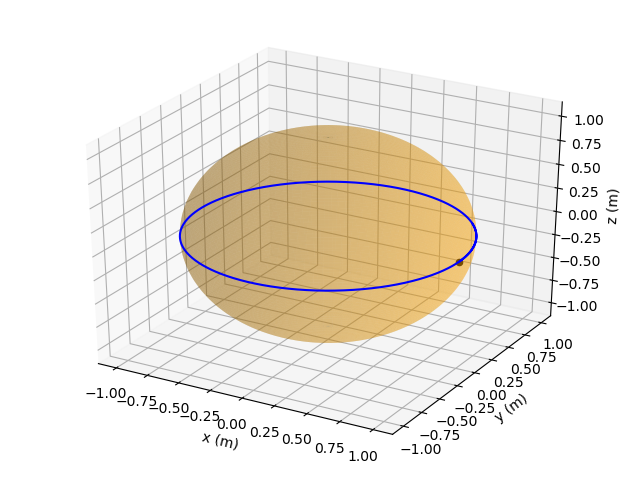}
  \centering
  \includegraphics[width=.49\linewidth]{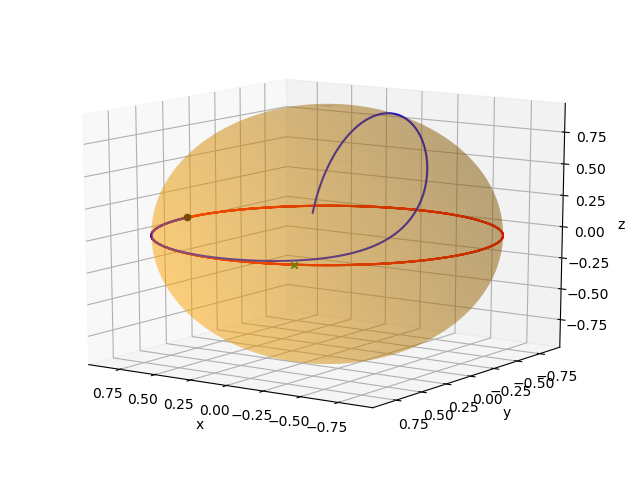}
\caption{Left: Geodesic on the Sphere. Black dot denotes the initial point. Right: Trajectory for the obstacle avoidance problem. Green cross denotes the obstacle.}
\label{fig1}
\end{figure}
\end{enumerate}

\begin{enumerate}
\item[(2)]\textit{Obstacle along cubics in tension:} Finally we show how the method works with cubics in tension. We consider the cubic in tension trajectory, with the point-obstacle along the curve and we want to design a trajectory that avoid the point-obstacle. We consider initial conditions $v(0)=(0,4,-1)$,  $v'(0)=(0,-0.3,0.5)$, $v''(0)=(0,-1,2)$ and $\sigma=1$. If we set the parameter $\tau=1$ we obtain practically the same trajectory. The trajectories are extremely close in value at all points, and in particular, the red curve is at a distance of $2.54\times10^{-6}$ from the obstacle placed in the blue curve as it is show in Figure \ref{fig4}.
  
\begin{figure}[h!]
\begin{center}
\includegraphics[width=.7\linewidth]{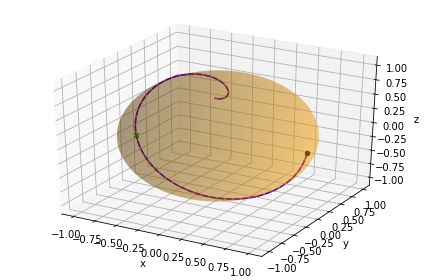}
\caption{Cubic in tension (blue, dashed line) with $\sigma=1$ and collision avoidance curve (red, solid line) with $\tau=1$.}
\label{fig4}\end{center}
\end{figure}



 However, we can increase $\tau$ and smoothly deform  the trajectory from the obstacle as $\tau$ increases. We show the comparison between the obstacle avoidance and the cubic in tension with the obstacle along it as $\tau$ increases in Figure \ref{fig5}.

\begin{figure}[h!]
\includegraphics[width=.49\linewidth]{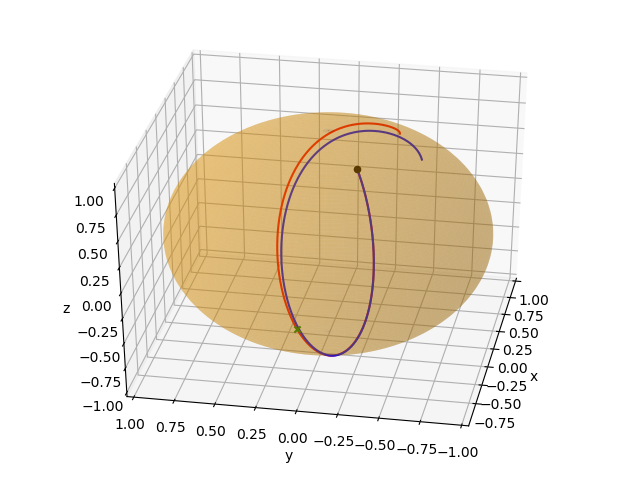}
  \centering
  \includegraphics[width=.49\linewidth]{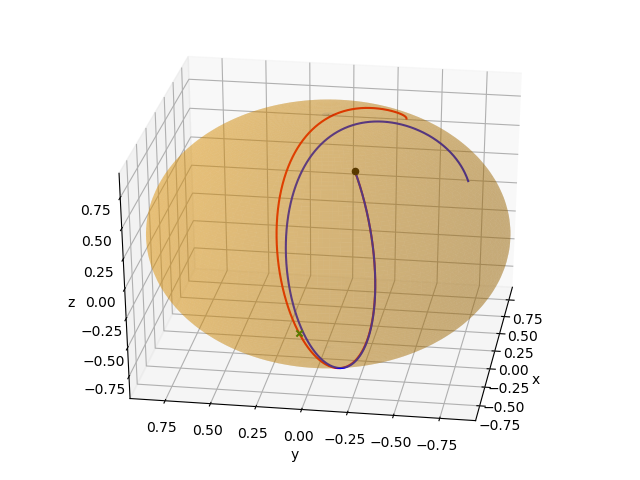}
 \centering
  \includegraphics[width=.49\linewidth]{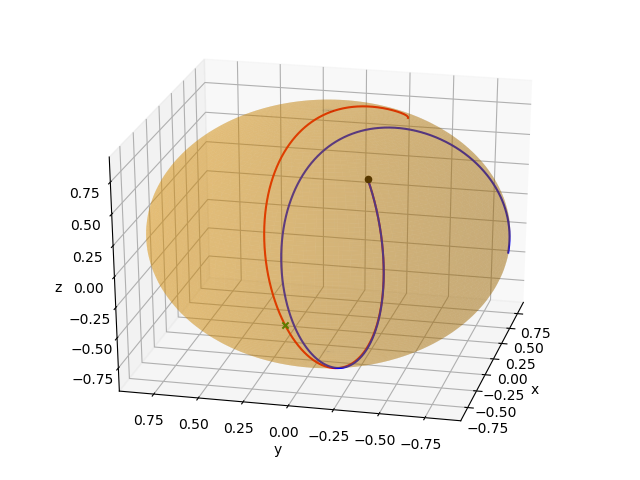}
\caption{Smooth deformation of the point-obstacle avoidance trajectory by increase $\tau$ from $\tau=50$, $\tau=200$ and $\tau=400$ with $\sigma=1$.}
\label{fig5}
\end{figure}

\end{enumerate}

\end{example}



\begin{thebibliography}{xx}
 \bibitem{ligia}L Abrunheiro, M Camarinha, J Clemente-Gallardo, JC Cuch\'{\i}, P Santos.  A general framework for quantum splines International Journal of Geometric Methods in Modern Physics 15 (09), 1850147, 2018.

\bibitem{mishal}
M. Assif, R. Banavar, A. Bloch, M. Camarinha, L Colombo.  Variational collision avoidance problems on Riemannian manifolds. in Proceedings of the IEEE International
Conference on Decision and Control, 2018, pp. 2791-2796.
  \bibitem{Bl} A. Bloch, J. Baillieul, P. E. Crouch, J. E. Marsden, D. Zenkov, Nonholonomic Mechanics and Control.  New York, NY: Springer-Verlag, 2nd ed. 2015.

\bibitem{BlCaCoCDC} A. Bloch, M. Camarinha, L. Colombo. Variational obstacle avoidance
on Riemannian manifolds. in Proceedings of the IEEE International
Conference on Decision and Control, 2017, pp. 146-150.
\bibitem{BlCaCoIJC} A. Bloch, M. Camarinha and L. J. Colombo. Dynamic interpolation for obstacle avoidance on Riemannian manifolds. International Journal of Control, pages 1-22, doi:10.1080/00207179.2019.1603400, \url{https://doi.org/10.1080/00207179.2019.1603400}.
Preprint available at. arXiv:1809.03168 [math.OC]
\bibitem{BlCr} A. Bloch and P. Crouch. On the equivalence of higher order variational
problems and optimal control problems. in Proceedings of the IEEE
International Conference on Decision and Control, Kobe, Japan, 1996,
pp. 1648-1653.
\bibitem{Boothby} W.  M.  Boothby, An  Introduction  to  Differentiable  Manifolds  and Riemannian Geometry. Orlando,  FL: Academic  Press Inc., 1975.

\bibitem{bookBullo}  F. Bullo and A. D. Lewis, Geometric Control of Mechanical Systems. Springer-Verlag,  2004.

\bibitem{MargaridaThesis} M. Camarinha. The geometry of cubic polynomials in Riemannian manifolds.
Ph.D. thesis, Univ. de Coimbra, 1996.

\bibitem{CoFeMdD}  L. Colombo, S. Ferraro, D. Martín de Diego. \textit{Geometric integrators for higher-order variational systems and their application to optimal control}. J. Nonlinear Sci. 26 (2016), no. 6, 1615-1650.
\bibitem{CoMdM} L. Colombo and D. Mart\'in de Diego. Higher-order variational problems on Lie groups and optimal control applications. J. Geom. Mech.
6 (2014), no. 4, 451-478.
\bibitem{Cophd}  L. Colombo. Geometric and numerical methods for optimal control of
mechanical systems. PhD thesis, Instituto de Ciencias Matem\'aticas,
ICMAT (CSICUAM-UCM-UC3M), 2014.

\bibitem{CroSil:95} P. Crouch and F. Silva Leite, The dynamic interpolation problem: on Riemannian manifolds, Lie groups, and symmetric spaces,  {\em J. Dynam. Control Systems} 1 (1995), no. 2, 177--202.

\bibitem{Giambo} R. Giamb\'o,
F. Giannoni, P. Piccione. ``An analytical theory for Riemannian cubic polynomials". IMA J Math Control Inform 19:445-460, 2002.

\bibitem{Helgason} S. Helgason, Differential geometry, Lie groups, and symmetric spaces, Pure and
Applied Mathematics, no. 80, Academic Press, Oxford, 1978.
\bibitem{Hinkle} J. Hinkle, P.T. Fletcher, S. Joshi. Intrinsic polynomials for regression on Riemannian
manifolds. Journal of Mathematical Imaging and Vision 50 (1-2), 32-52, 2014.
\bibitem{Hong}
Y. Hong, R. Kwitt, N. Singh, N. Vasconcelos, and M. Niethammer. Parametric Regression on the Grassmannian IEEE Transactions on Pattern Analysis and Machine Intelligence, 2016.

\bibitem{BlochHussein}
I. Hussein  and A. Bloch.
\textit{Dynamic Coverage Optimal Control for Multiple Spacecraft Interferometric Imaging}. Journal of Dynamical and Control Systems, Vol. 13, Issue 1, pp 69-93, 2007

\bibitem{Jackson} J. Jackson, \textit{Dynamic interpolation and application to flight control}. PhD Thesis.Arizona State Univ., 1990.
\bibitem{Karcher77}H. Karcher. Riemannian center of mass and mollifier smoothing
Comm. Pure Appl. Math., 30: 509-541, 1977.
\bibitem{Khatib1986} O. Khatib. \textit{Real-time obstacle avoidance for manipulators and mobile robots}. Int. J. of Robotics Research, vol 5, n1, 90--98,1986.

\bibitem{K88}
D. Koditschek. \textit{Robot planning and control via potential functions}
Robotics Review. MIT Press, Cambridge, MA. 1992.
\bibitem{Koditschek1990}
D. Koditschek and E. Rimon. \textit{Robot navigation functions on manifolds with boundary}
Adv. in Appl. Math., Vol11, n4, 412--442, 1990.

\bibitem{Milnor} J. Milnor, Morse Theory.  Princeton, NJ: Princeton Univ. Press, 2002.

\bibitem{Noa:89} L. Noakes, G. Heinzinger and B. Paden, Cubic Splines on
Curved Spaces, {\em IMA Journal of Math. Control \& Inf.} 6,
(1989), 465--473.
\bibitem{Nomizu} K. Nomizu, \textit{Invariant affine connections on homogeneous spaces}. Amer. J. Math 76, pp. 33-65, 1954.
\bibitem{Oneill} B. O' Neill. Submersions and geodesics. Duke Math. J. 34, 363-373, 29, 1967.
\bibitem{SCC00} F. Silva Leite, M. Camarinha and P. Crouch, Elastic curves as solutions of Riemannian and sub-Riemannian control problems {\em Math. Control Signals Systems} 13, no. 2, 140--155, (2000).
\bibitem{ZN} E. Zhang, L. Noakes. Left Lie reduction for curves in homogeneous spaces. Adv Comput Math.44 (5), 1673-1686, 2018.
\bibitem{ZN1}E Zhang, L Noakes. Relative geodesics in bi-invariant Lie groups
Proceedings of the Royal Society A: Mathematical, Physical and Engineering Sciences, 473, 20160619, 2017.
\end{thebibliography}
\end{document}